\documentclass[preprint,onecolumn,superscriptaddress,tightenlines,nofootinbib,11pt]{article}
\pdfoutput=1 
\usepackage{graphicx}
\usepackage{epsfig}
\usepackage{amsmath,latexsym,amssymb,hyperref,enumerate}
\usepackage{amsfonts}
\usepackage{url}
\usepackage{color}
\usepackage{cancel}
\usepackage{cite}
\usepackage[font={sf}]{caption}
\hypersetup{colorlinks,citecolor= nicegreen,linkcolor= nicered}
\definecolor{nicered}{rgb}{0.7,0.1,0.1}
\definecolor{nicegreen}{rgb}{0.1,0.5,0.1}
\newcommand{\bg}[1]{\mbox{\boldmath{$#1$}}}
\begin{document}

\normalsize
\vspace{-0.5cm}
\begin{flushright}
CALT-TH-2016-005\\
\vspace{1cm}
\end{flushright}

\vspace*{1.0cm}
\centerline{\bf\Large\textcolor[rgb]{0,0,0}{Effects of Bound States on Dark Matter Annihilation}}
\vspace*{1.2cm}

\centerline{\bf 
{Haipeng~An \footnote{\tt anhp@caltech.edu}}, \ \ 
{Mark~B.~Wise \footnote{\tt wise@theory.caltech.edu}}, \ \  
{Yue~Zhang \footnote{\tt yuezhang@theory.caltech.edu}}
}

\vspace*{1.0cm}

\centerline{\em Walter Burke Institute for Theoretical Physics,}
\centerline{\em California Institute of Technology, Pasadena, CA 91125}

\vspace{1.5cm}

\parbox{16.5cm}{
{\sc Abstract:} 
We study the impact of bound state formation on dark matter annihilation rates in models where dark matter interacts via a light mediator, the dark photon. We derive the general cross section for radiative capture into all possible bound states, and point out its non-trivial dependence on the dark matter velocity and the dark photon mass. For indirect detection, our result shows that dark matter annihilation inside bound states can play an important role in enhancing  signal rates over the rate for direct dark matter annihilation with Sommerfeld enhancement. The effects are strongest for large dark gauge coupling and when the dark photon mass is smaller than the typical momentum of dark matter in the galaxy. As an example, we show that for thermal dark matter the Fermi gamma ray constraint is substantially increased once bound state effects are taken into account. We also find that bound state effects are not important for dark matter annihilation during the freeze out and recombination epochs.
}

\newpage

\section{Introduction}

It is very likely that dark matter (DM)  requires  degrees of freedom, that are not in the standard model (SM) for its explanation.  As its name implies the electromagnetic interactions of DM must be small. A convenient way to realize this is to suppose that the DM is not charged under the SM gauge group. A simple model of this type that doesn't have any fine tunings, beyond the usual ones to keep the cosmological constant and Higgs mass small, is to have the DM be a Dirac fermion  coupled to a new massive $U(1)_D$ gauge boson (the dark photon). Since the DM couples to a conserved current the new gauge boson can have a mass term in the Lagrange density. This model  adds to the SM seven new dark degrees of freedom: two spin components for both the DM and anti-DM particles, and the three spin states for the massive dark photon.  The model has three additional parameters, a dark fine structure constant, $\alpha_D=g_D^2/(4 \pi)$, the DM mass $m_D$ and the dark photon mass $m_V$. In addition there is one dimensionless renormalizable coupling $\kappa$ that characterizes the kinetic mixing of the hypercharge $U(1)_Y$ and $U(1)_D$ kinetic terms. It is only through gravity and this kinetic mixing that SM degrees of freedom communicate with those in the dark sector. 

In this paper we will assume thermal DM {\it i.e.}, at early times when the universe is at a very high temperature  the DM sector is in thermal equilibrium, and moreover has the same temperature as the SM. As the universe evolves it cools and when the temperature drops below the DM mass, DM and anti-DM particles start to annihilate, eventually freezing out at $T \sim m_D/30$. In this scenario achieving the correct the DM density relates the dark fine structure constant to the DM mass, roughly $\alpha_D \sim 0.02\,(m_D/ 1 \,{\rm TeV})$. 
For DM much heavier than TeV, one needs to take into account of the Sommerfeld effect during freeze out and $\alpha_D$ is somewhat lower than the value predicted by the above relation. For example, $\alpha_D=0.2$ for $m_D=16.7$\,TeV.

The light mediator scenario with $m_D \gg m_V$ has been studied for a variety of reasons. In the same region of parameter space there are  indirect detection signals that are the topic of study in this paper. For $\alpha_Dm_D/(2 m_V)>0.84$, two body DM-anti-DM bound states exist. They are the analog of positronium bound states in electromagnetism. It is the impact of these darkonium bound states on the rate for DM-anti-DM annihilation in the galaxy today that we focus on. 

\begin{figure}[t]
\centerline{\includegraphics[width=0.6\columnwidth]{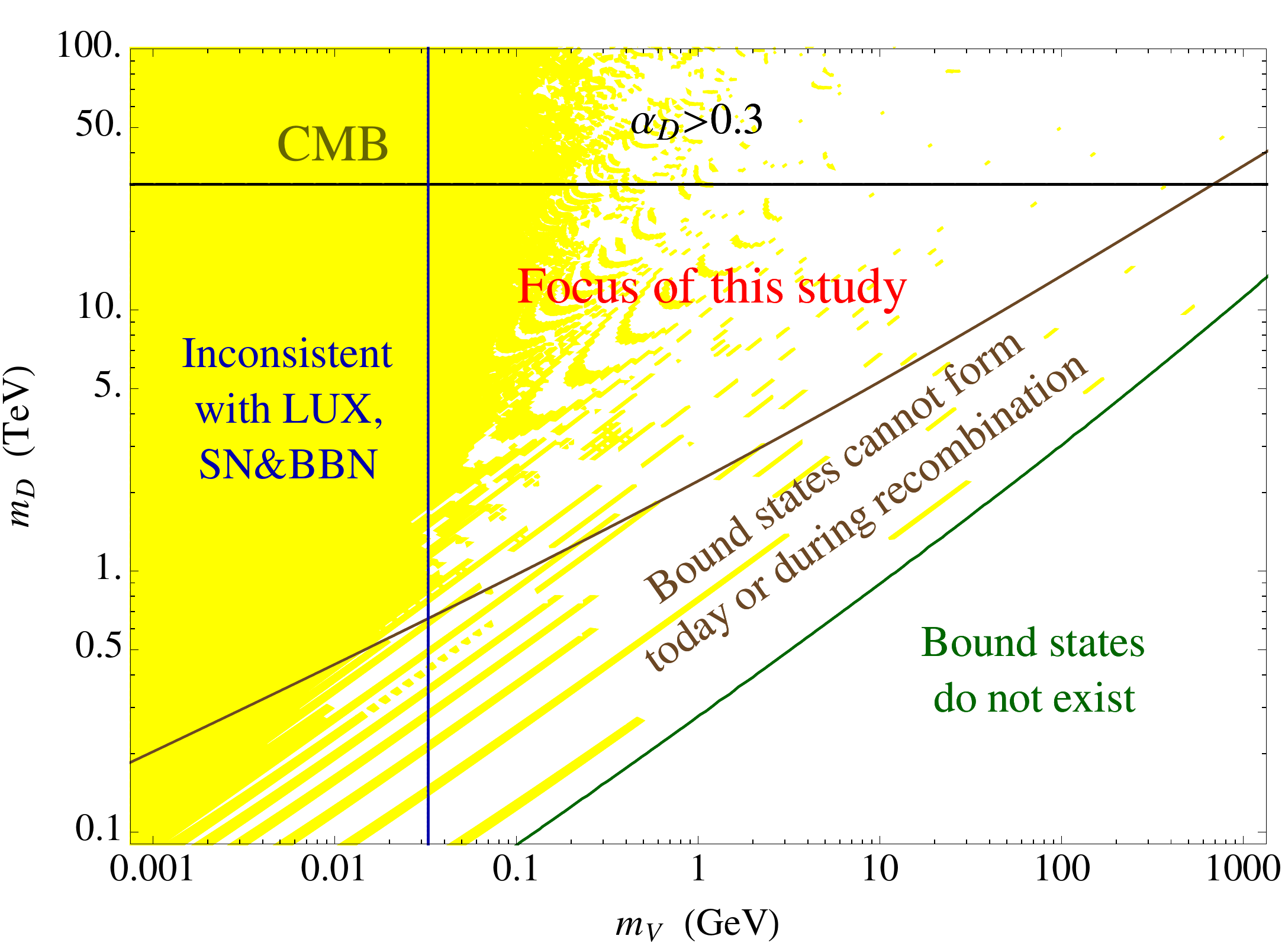}}
\caption{
The parameter space relevant to this study (marked by ``Focus of this study''), where $m_D$ is the dark matter mass and $m_V$ is the dark photon mass. 
The value of the dark fine structure constant $\alpha_D = g_D^2/(4\pi)$ is chosen to give the correct relic abundance for dark matter.
We do not consider $\alpha_D<0.3$ (above the black curve) where the next-to-leading order corrections would be large.
Dark matter bound states do not exist below the green curve. 
Between the brown and green curves, dark matter bound states exist but the dark photon is too heavy for the bound states to be produced with the low kinetic energy such as in the galaxy today or during recombination.
The region to the left of the blue curve is excluded by direct detection, the supernova constraints and demanding that dark photons decay before BBN, see also
Fig.~\ref{DecayFast}. The yellow region is excluded by CMB constraints on dark matter annihilation during recombination, see section~\ref{cmb}.
}\label{parameterspace}
\end{figure}

The region of parameter space in the $m_V-m_D$ plane where bound states exist and are potentially important for DM cosmology is shown in Fig.~\ref{parameterspace}.  We assume these darkonium states are non-relativistic and so we restrict our attention to the region of parameter space where $\alpha_D <0.3$ or equivalently (for thermal DM $m_D<30\, {\rm TeV}$). Very small dark photon masses, $m_V < 30\, {\rm MeV}$ are inconsistent with direct detection, the supernova constraints and the requirement that dark photons decay away before big bang nucleosynthesis (BBN). The region of Fig.~\ref{parameterspace} below the green line does not have darkonium states and the region between the brown and green lines has darkonium states but the mass of the dark photon is too large for these bound states to be produced with the low kinetic energy for the DM during recombination or today. Thus, the region of parameter space for our study of the impact of bound states on indirect detection signals is the triangular region marked  as the ``Focus of this study'' in Fig.~\ref{parameterspace}. 

The main finding of our work is that the bound state effects are important for DM indirect detection when $m_V/m_D\lesssim10^{-3}$ and $\alpha_D \gtrsim 0.1$. The goal of this paper is not to provide a comprehensive list of all the indirect detection constraints but rather to highlight the important role bound state formation plays in this region of parameter space. To this end we focus on the photon spectrum resulting from DM near the center of our galaxy annihilating either directly or through bound states.

In section \ref{generalformula}, we derive the general cross section for dark matter bound state formation via radiation of an on-shell dark photon and its dependence on the dark matter velocity and the dark photon mass. In section \ref{ID}, we apply our results to calculate the indirect detection constraints on dark matter annihilation and discuss the importance of bound state effects. We discuss the bound state effects on dark matter freeze out and on the cosmic microwave background (CMB) in sections \ref{TRD} and \ref{cmb}, and conclude in section \ref{finale}.

\section{Bound State Formation Cross Section}\label{generalformula}

The Lagrangian for the dark sector is
\begin{eqnarray}\label{Lvector}
\mathcal{L} &=& \mathcal{L}_{\rm SM} + \bar\chi i\gamma^\mu(\partial_\mu - i g_D V_\mu) \chi - m_D \bar \chi \chi -\frac{1}{4} V_{\mu\nu} V^{\mu\nu} + \frac{1}{2} m^2_{V} V_\mu V^\mu - \frac{\kappa}{2 \cos\theta_w}B_{\mu\nu} V^{\mu\nu} \ ,
\end{eqnarray}
where $B_{\mu\nu}$ is hyper-charge field strength tensor and $\theta_w$ is the Weinberg angle. Hence $\kappa$ is the kinetic mixing between the photon and the vector field $V$.

The interaction Hamiltonian for radiating one dark photon $V$ can be separated into transverse and longitudinal mode parts.
In the center-of-mass frame, the Hamiltonian for radiating transverse $V$'s is
\begin{eqnarray}
H_{int}^T = \left({g_D {\bf k} \over \mu}\right) \left[ {\bf A}_T\left( \frac{\bf r}{2} \right) + {\bf A}_T\left( -\frac{\bf r}{2} \right) \right] \ ,
\end{eqnarray}
where ${\bf r}$ is the relative coordinate of $\chi$ and $\bar \chi$,  ${\bf k}$ is the relative DM momentum and $\mu=m_D/2$ is the DM reduced mass. For the transverse modes, the polarization vectors satisfy $\epsilon_i^+(q) \epsilon_j^{+*}(q) + \epsilon_i^-(q) \epsilon_j^{-*}(q)=\delta_{ij}-q_iq_j/|{\bf q}|^2$, and ${\bf q}$ is the three momentum of the radiated dark photon. For radiating a longitudinal $V$, using current conservation $q^\mu J_\mu = 0$, the interaction Hamiltonian can be written as
\begin{eqnarray}
H_{int}^L = \left( {g_D m_V\over |{\bf q}|}\right)\left[ \phi_L\left( \frac{\bf r}{2} \right) - \phi_L\left( -\frac{\bf r}{2} \right) \right] \ .
\end{eqnarray}
Effectively it is equivalent to radiating a scalar particle.

For bound state formation, $|{\bf q}| \sim \alpha_D^2 \mu$ and $|{\bf r}| \sim 1/(\alpha_D \mu)$, where $\alpha_D = g_D^2/(4\pi)$. 
Thus we use the dipole approximation and only keep the leading terms in ${\bf q} \cdot {\bf r}$. Then the matrix elements for the free-bound transitions with transverse and longitudinal $V$ radiation are
\begin{eqnarray}
\mathcal{M}_T = g_D \int d^3{\bf r}\ \Psi_f^*({\bf r})(E_i - E_f) {\bf r} \cdot {\bg \epsilon}_T  \Psi_i({\bf r}) \ , \ \ \ 
\mathcal{M}_L = -ig_D \int d^3{\bf r}\ \Psi_f^*({\bf r})   m_V \frac{{\bf r} \cdot {\bf q}}{|{\bf q}|}\Psi_i({\bf r}) \ .
\end{eqnarray}

The total cross section for bound state formation from a scattering state with momentum ${\bf k}$ at infinity is
\begin{eqnarray}\label{GeneralizedK}
(\sigma v)_{\rm B} = \frac{\alpha_D}{3 \pi} \sum_{n, \ell} \left( \omega_{n\ell}^2 + \frac{1}{2} m_V^2 \right) \sqrt{\omega_{n\ell}^2 - m_V^2} \left[ \ell \left|\int dr r^3 R_{n\ell}(r) R_{k \ell-1} \right|^2 +(\ell+1) \left|\int dr r^3 R_{n\ell}(r) R_{k \ell+1} \right|^2 \right] \ ,
\end{eqnarray}
where $v$ is the relative velocity and $\omega_{n\ell} = E_{n\ell} + k^2/(2\mu)$ is the sum of the binding energy of the $(n\ell)$'th bound state level and the kinetic energy of incoming state.  For a Yukawa potential, the binding energy in general depends on both $n$ and $\ell$.

\medskip
A couple of remarks relevent for the evaluation of Eq.~(\ref{GeneralizedK}) are:
\begin{itemize}
\item The sum over $n, \ell$ includes all the energy level satisfying $\omega_{n\ell}>m_V$. For low velocity DM with $k^2/(2\mu) \ll m_V$, this amounts to $E_{n\ell}>m_V$. As a rough estimate we can use $E_{n\ell} \sim \alpha_D^2 \mu/(2n^2)$ for the binding energy. Then if $E_{n\ell}>m_V$, the ratio of the bound state size to the screening length of the Yukawa potential, $a_n m_V \sim n m_V/(\alpha_D \mu)$  is less than $\alpha_D/(2n)\ll1$. In other words those bound states, that are deep enough to emit an on-shell $V$, in their formation,  are all much smaller than $1/m_V$. Therefore, from now on, we will  approximate  the relevant bound states as Coulomb bound states, with energies $E_{n\ell} = E_n$, $\omega_{n\ell} = \omega_n$, that are $\ell$ independent.

\item The bound and scattering wavefuctions that solve the Schr\"odinger equation are written as
\begin{eqnarray}
\psi_n({\bf r}) = \sum_{\ell m} R_{n\ell}(r) Y_{\ell m}(\hat r), \hspace{1cm} \psi_k({\bf r}) = \sum_{\ell m} R_{k\ell}(r) Y_{\ell m}(\hat r) Y^*_{\ell m}(\hat k) \ .
\end{eqnarray}
Using the above approximation, the radial Coulomb wavefuctions for bound states are
\begin{eqnarray}
R_{n \ell}(r) = \frac{2}{n^{\ell+2} (2\ell+1)!} \sqrt{\frac{(n+\ell)!}{(n-\ell-1)!}}\ \frac{(2r)^\ell}{a_0^{\ell+3}} e^{-\left( r/{n a_0}\right)} F_1 \left( 1+\ell -n, 2+2\ell, \frac{2 r}{n a_0} \right) \ ,
\end{eqnarray}
where $a_0=1/(\alpha_D \mu)$ is the Bohr radius.

For the scattering state radial wave functions, we numerically solve the Schr\"odinger equation with a Yukawa potential and energy eigenvalue $E=k^2/(2\mu)$ using the ``shooting method''. For the $\ell$'th partial wave, define $R_{k\ell}(r) = r^{\ell-1} \phi(r)$, the Schr\"odinger equation for $\phi(r)$ is
\begin{eqnarray}\label{eq8}
\phi''(r) + \frac{2\ell}{r} \phi'(r) + \left(k^2 + \frac{2 \alpha \mu e^{- m_V r}}{r} - \frac{2 \ell}{r^2}\right) \phi(r) =0 \ .
\end{eqnarray}
The boundary condition at $r=0$ is $\phi(0)=0$ and $\phi'(0)=c$. 
Because Eq.~(\ref{eq8}) is a linear equation,
the overall normalization of $R$ is proportional to the parameter $c$, and we fix it by requiring that $R_{k\ell}$ matches to the $\ell$-th partial of a plane wave as $r \to \infty$, {\it i.e.},
\begin{eqnarray}
r R_{k\ell} (r\to \infty) \sim \frac{(4\pi)}{k} i^\ell \cos \left(kr - (\ell+1)\frac{\pi}{2} + \delta_{k\ell} \right) \ .
\end{eqnarray}
In the $m_V\to0$ limit, $R_{k\ell}$ is given by the Coulomb scattering wave function,
\begin{eqnarray}
R_{k \ell}(r) = \frac{4\pi e^{\frac{\pi}{2 k a_0}}\left| \Gamma(1+\ell -\frac{i}{k a_0}) \right|}{(2\ell+1)!} (2 k r)^\ell e^{- i k r}\,\! _1F_1 \left( 1+\ell + \frac{i}{ka_0}, 2+2\ell, 2ikr \right) \ .
\end{eqnarray}

\end{itemize}

\begin{figure}[t]
\centerline{\includegraphics[width=0.6\columnwidth]{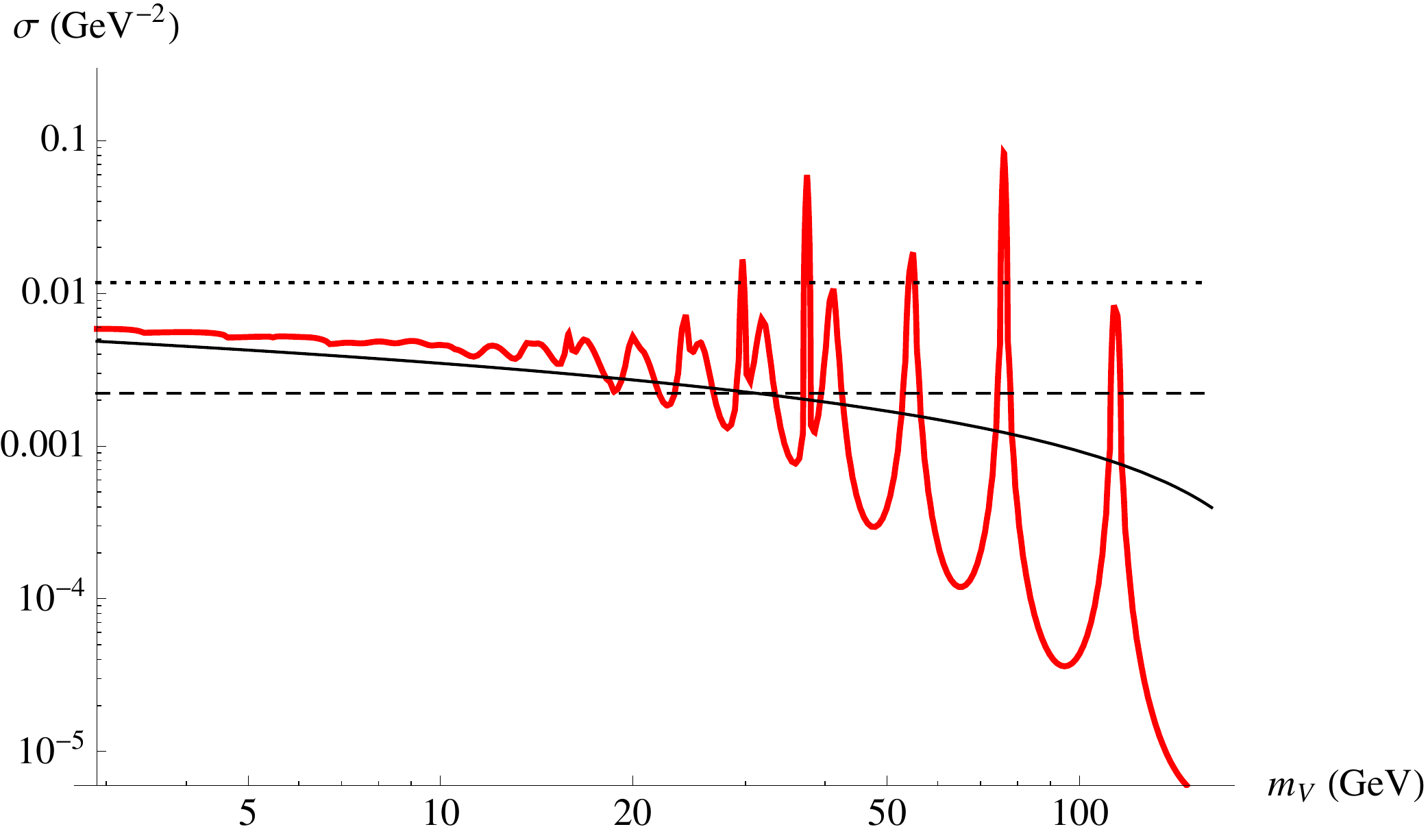}}
\caption{
Dark photon mass dependence in the bound state formation cross section in today's galaxy (thick red curve), 
calculated from the general formula, Eq.~(\ref{GeneralizedK}). 
We have fixed the other parameters to be $m_D=16.7\,{\rm TeV}$, $\alpha_D=0.2$ and the velocity $v=10^{-3}$. 
The solid black curve is obtained from the modified Kramers formula Eq.~(\ref{MdofiedKramers}), which does not capture the resonance effects.
The horizontal dotted line is the original Kramers formula in the Coulomb limit ($m_V\to0$), Eq.~(\ref{Kramers}), 
and the dashed line corresponds to only ground state formation with massless dark photon with $n=1$.
}\label{mVdependence}
\end{figure}

\noindent We first discuss the $m_V$ dependence of the bound state formation cross section in Eq.~(\ref{GeneralizedK}). In the Coulomb limit $m_V\to 0$, it is the Kramers formula for the recombination cross section~\cite{kramers}. 
For $\alpha_D \gg k/\mu \equiv v$, the leading terms of the Kramers formula for the cross section are~\cite{Katkov}
\begin{eqnarray}\label{Kramers}
\sigma_{\rm B} = \frac{32\pi}{3\sqrt{3}} \frac{\alpha_D^3}{\mu^2 v^2} \left[ \ln \left( \frac{\alpha_D}{v} \right)+ 0.16 + \mathcal{O}(v/\alpha) \right] \ .
\end{eqnarray}
The logarithmic factor arises from the sum over $n$ in Eq.~(\ref{GeneralizedK}). For given level $n\gg1$~\cite{bethebook},
\begin{eqnarray}\label{KramersN}
(\sigma_{\rm B})_n \simeq \frac{32\pi}{3\sqrt{3}} \frac{\alpha_D}{\mu^2} \frac{E_0^2}{ \left(\mu v^2/2 \right)\left(\mu v^2/2 + {E_0}/{n^2} \right) n^3} \ .
\end{eqnarray}
Recall $E_0$ is the binding energy of the Coulomb bound state, $E_0=\alpha_D^2 \mu/2$. The condition $\alpha_D > v$ implies $E_0 > \mu v^2/2$. It is important to note that $\sigma_n \sim 1/n$ for small $n$ and $\sigma_n \sim 1/n^3$ for large $n$. The transition between these two behavious occurs around $n_{\rm trans} \sim \alpha_D/v$. The sum of $\sigma_n$ from $n=1$ to $n_{\rm trans}$ results in the logarithmic factor $\ln(\alpha_D/v)$ in Eq.~(\ref{Kramers}). 
\footnote{We have verified the Kramers formulae~(\ref{Kramers}) and (\ref{KramersN}) numerically.  
The calculations taking into account only the capture into ground state~\cite{Pospelov:2008jd, MarchRussell:2008tu} would miss the logarithmic factor and underestimate the cross section, for the cases $\alpha_D\gg v$ and/or $\alpha_D \gg \sqrt{m_V/m_D}$.
We also note that Refs.~\cite{Feng:2009mn, vonHarling:2014kha} tried to sum over all the energy levels ($n\geq1$) but concluded it yields a factor of $\pi^2/6$ compared to the ground state case.
However, they used the recombination cross section Eq.~(75.6) in~\cite{bethebook} that only applies for $n=1$ and also underestimated the enhancement. 
Moreover, these previous works have assumed massless dark photon limit which would neglect the nontrivial $m_V$ dependence as shown in Fig.~\ref{mVdependence}.
}

For $m_V\neq0$ the situation is more complicated.  For non-zero mediator mass $m_V$, we solved for the cross section numerically using the approach described above.
As $m_V$ grows, the first effect it has is to change the upper limit of the sum over $n$. The largest $n_{\rm max}$ corresponds to the highest energy level that has a large enough binding energy that allows $V$ to be produced on-shell. When we are still within the Coulomb limit, $n_{\rm max} = \alpha_D \sqrt{\frac{\mu}{2 m_V}}$. When $n_{\rm max}<n_{\rm trans}$, the Kramers cross section is modified to
\begin{eqnarray}\label{MdofiedKramers}
\sigma_{\rm B} \simeq \frac{32\pi}{3\sqrt{3}} \frac{\alpha_D^3}{\mu^2 v^2} \ln \left( \alpha_D \sqrt{\frac{\mu}{2 m_V}} \right) \ .
\end{eqnarray}
The condition that $n_{\rm max}<n_{\rm trans}$ is equivalent to $ \mu v^2/2 <m_V$ and  we find numerically that Eq.~(\ref{MdofiedKramers}) is valid in the range $\frac{1}{2} \mu v^2 < m_V< \mu v$. 
In the region $m_V  >  \mu v$, the cross section is resonantly enhanced when the kinetic energy of the incoming state is nearly degenerate with a resonance state of the Yukawa potential. These features are shown in Fig.~\ref{mVdependence}. 

\begin{figure}[t]
\centerline{\includegraphics[width=0.6\columnwidth]{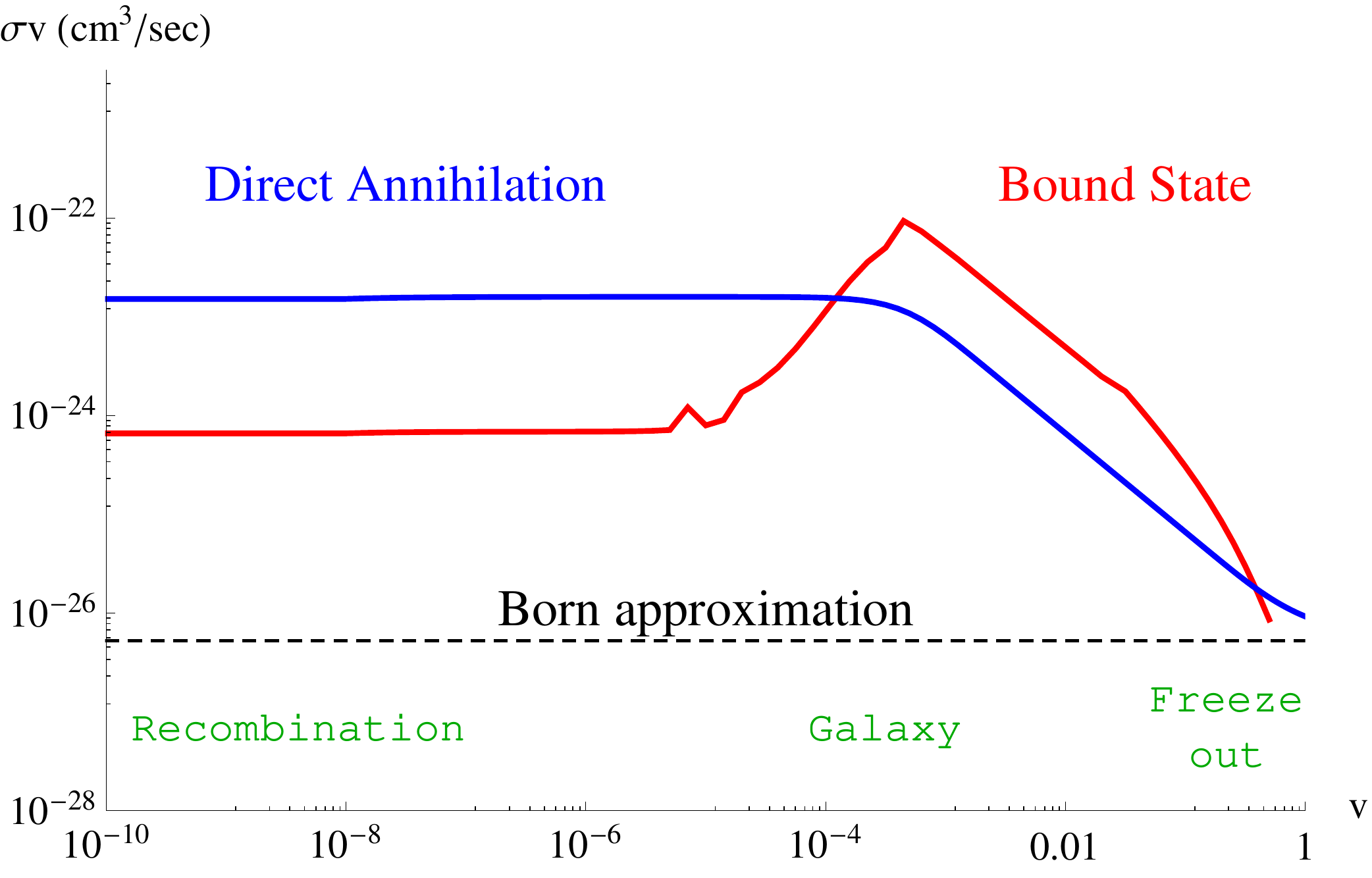}}
\caption{
Dark matter bound state formation cross section for various dark matter velocities (thick red curve), calculated from the general formula, Eq.~(\ref{GeneralizedK}). 
We have fixed the other parameters, $m_D=16.7\,{\rm TeV}$, $\alpha_D=0.2$ and $m_V=10\,{\rm GeV}$. 
For comparison, we also show the direct annihilation cross section in the Born approximation (dashed black) and the one with Sommerfeld enhancement (thick blue curve).
The bound state formation cross section plays the most important role for dark matter indirect detection, if $m_V/m_D < v < \alpha_D$.
}\label{velocitydependence}
\end{figure}

Next, we discuss the velocity dependence of the bound state formation cross section, and highlight the comparison with the Sommerfeld enhanced cross section of direct annihilation, often used for computing DM annihilation in the literature~\cite{Cirelli:2008pk, ArkaniHamed:2008qn, Pospelov:2008jd, Fox:2008kb}. This comparison is shown in Fig.~\ref{velocitydependence} for $m_D=16.7\,{\rm TeV}$, $\alpha_D=0.2$ and $m_V=10\,{\rm GeV}$.
 We find that in the region $m_V/\mu < v < \sqrt{2 m_V/\mu}$  the bound state formation cross section is consistently larger than the direct annihilation cross section with Sommerfeld enhancement (labelled with subscript A). For this range of relative velocities, the two cross sections can be approximated as
\begin{eqnarray}
(\sigma v)_{\rm B} \sim \frac{32\pi \alpha_D^3}{3\sqrt{3} \mu^2 v} \ln \left( \alpha_D \sqrt{\frac{\mu}{2 m_V}}\right) , \hspace{1cm} 
(\sigma v)_{\rm A} \sim \frac{\pi^2 \alpha_D^3}{2 \mu^2 v} \ .
\end{eqnarray}
The direct annihilation cross section $(\sigma v)_{\rm A}$ is obtained by enhancing the Born level cross section by the $s$-wave Sommerfeld factor, defined as ${|R_{k, \ell=0}|^2}/({4\pi})^2$ in our convention.

The ratio of the two above cross sections is
\begin{eqnarray}
\frac{\rm Bound\ state\ formation\ rate}{\rm Direct\ Annihilation\ rate} = \frac{(\sigma v)_{\rm B}}{(\sigma v)_{\rm A}} = \frac{64}{3\sqrt{3} \pi}\ln \left( \alpha_D \sqrt{\frac{\mu}{2 m_V}}\right) \ .
\end{eqnarray}
With the parameters used in Fig.~\ref{velocitydependence}, the bound state formation cross section can be  larger by more than one order of magnitude over the Sommerfeld enhanced annihilation cross section.

For $v>\sqrt{2 m_V/\mu}$, the bound state production cross section is  given by the Kramers formula in Eq.~(\ref{Kramers}) and the the logarithmic enhancement factor is suppressed compared with the region of velocity we have just discussed.  This is the region to the right in Fig.~\ref{velocitydependence}. Eventually, at $v\sim \alpha_D$, the argument of the log factor is $\sim 1$ and the Kramers and Sommerfeld cross sections become comparable to each other. 

For $v<m_V/\mu$, the kinetic energy of the incoming state $\frac{1}{2} \mu v^2$ becomes smaller than the height of bump of the Yukawa potential barrier $V_{\rm barrier} \sim {\ell(\ell+1)m_V^2}/{\mu}$  for the $\ell\neq 0$ partial wave, located near $r \sim 1/m_V$.
As $v$ decreases, it becomes increasingly more difficult for these partial wave states to penetrate through the barrier to find the bound state wavefunction. 
In this region, the contributions to bound state formation cross section from all partial waves with $l \ne 0$ are suppressed. 
Eventually, at very tiny $v$, only the $ks \to np, (n\geq2)$ transitions can happen.  In this region the bound state formation cross section is smaller  than the Sommerfeld enhanced annihilation cross section.

The above velocity dependence can have important impact on indirect detection of DM annihilation in the Milky way galaxy, where the DM velocity is $\sim 10^{-3}$. We find for $m_V/\mu< 10^{-3}$ and $\alpha_D > 0.1$ (corresponding to multi-TeV scale thermal DM), it is much more likely for two DM particles to form a bound state than directly annihilate. 

After a bound state is formed, it could either annihilate decaying to dark photon $V$'s or de-excite to a lower state. The annihilation decay rate for the $n{ \ell}$ bound state goes as,
\begin{eqnarray}
\Gamma_{n,s,\ell \to V's} \sim \left({\alpha_D \over n} \right)^{2 \ell+3} \alpha_D^{(5-C)/2} \mu \ ,
\end{eqnarray}
where $s= 0, 1$ is the total spin angular momentum of the bound state, $n$ is the principal quantum number, $\ell$ is the orbital angular momentum, and $C=(-1)^{\ell+S}$ is the charge conjugation. For $C=1$, the bound state decays into $2V$'s, while for $C=-1$, it  decays into $3V$'s due to the Furry's theorem in the dark sector. The $\ell$ dependence arises because the annihilation decay amplitude is proportional to the $\ell$-th derivative of the zero point wavefunction at the origin, $({d^\ell}/{d r^\ell}) R_{n\ell}(0)$. Each derivative yields a power of $\alpha_D$. For smaller $n$ (and hence $\ell$) the time scale for darkonium annihilation to dark gauge bosons  is extremely short compared with the  the age of our galaxy which in units of inverse ${\rm GeV}$ is, $\tau_g\sim 0.62 \times 10^{42}\,{\rm GeV}^{-1}$. But for larger principal quantum numbers $n$ and values of $\ell$ that is not  the case. As an explicit example, we consider the parameters $m_D=16.7\, {\rm TeV}$, $\alpha_D=0.2$ and $m_V=1\,{\rm GeV}$. For these parameters $n_{\rm max}=12$ and for this value of the principal quantum number, $\ell=n_{max}-1=11$ and $C=-1$ the bound state annihilation decay has a lifetime associated with it that is about one order of magnitude larger than the age of our galaxy. 

Darkonium states with larger $n$ and $\ell$  produced in our galaxy do disappear but the route is through de-excitation to lower values of $n$ and $\ell$ and then annihilation to energetic $V$'s with $E_V \simeq m_D/2$.  For simplicity we consider  the case where the transition is dark electric dipole to either a real or if that is kinematically not allowed virtual $V$.  When this can occur via a real dark $V$  the rate is very rapid.  For  de-excitation through a virtual $V$ we estimate the rate for the bound state transition $n,\ell \to n-1,\ell\pm1$  to be
\begin{eqnarray}
\Gamma_{n \rightarrow n-1} \sim  \frac{ \kappa^2 \alpha \alpha_D^{13}}{n^{19}} \frac{\mu^5}{4 \pi^2m_V^4} \ .
\end{eqnarray}
It is convenient to introduce the ``partial lifetimes'', $\tau_{n \rightarrow n-1}=1/\Gamma_{n \rightarrow n-1}$. 
Assuming the transition to the ground state occurs changing $n$ by one unit at a time the total rate is
\begin{eqnarray}
\Gamma_{n\to1} \sim\left( {1{\Bigg/} {\sum_{i=2}^{n} \tau_{i \rightarrow i-1} }}\right) \sim 20 \frac{ \kappa^2 \alpha \alpha_D^{13}}{n^{20}} \frac{\mu^5}{4 \pi^2m_V^4} \ .
\end{eqnarray}

\begin{figure}[t]
\centerline{\includegraphics[width=0.65\columnwidth]{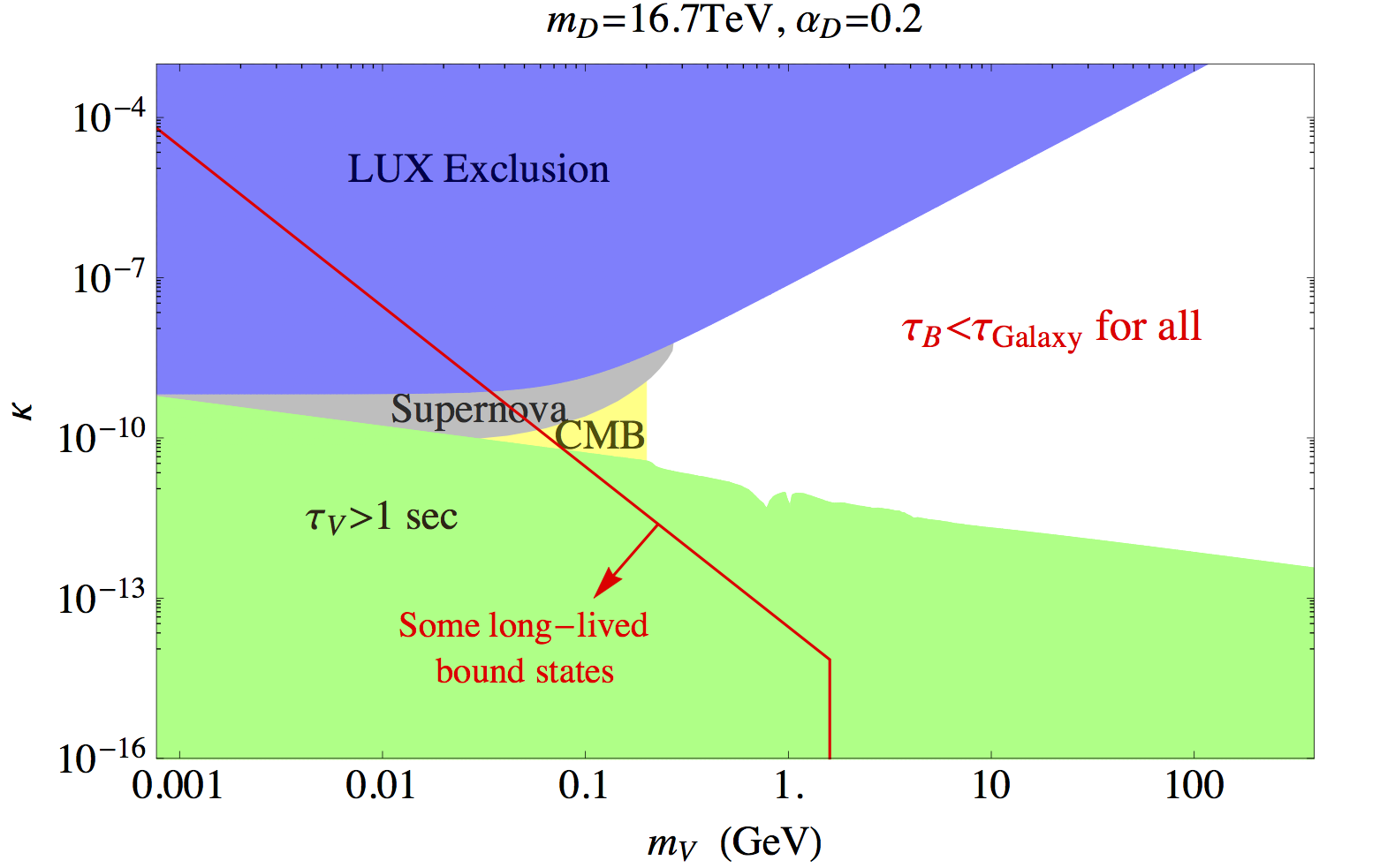}}
\caption{
Constraints on the $m_V - \kappa$ parameter space for fixed $m_D=16.7\,{\rm TeV}$ and $\alpha_D=0.2$.
The blue region is excluded by the LUX result in dark matter direct detection~\cite{Akerib:2015rjg}.
The gray region is excluded from the supernova cooling argument~\cite{Bjorken:2009mm, Dent:2012mx}.
In the green region, the dark photon lives longer than a second and would threaten the success of BBN assuming the dark and SM sectors had similar temperatures~\cite{Zhang:2015era, Kaplinghat:2013yxa}. In the allowed (white) region, all possible dark matter bound states have lifetime shorter than the age of the Milky Way galaxy.
}\label{DecayFast}
\end{figure}

We find in this case for the allowed values of $\kappa$, $m_D$ and $m_V$ that this de-excitation rate is shorter than the age of the galaxy. This is illustrated in Fig.~\ref{DecayFast} using the same parameters as before ({\it i.e.}, $m_D=16.7~ {\rm TeV}$, $\alpha_D=0.2$, $m_V=1{\rm GeV}$ and $n_{\rm max}=12$).  All the darkonium bound states produced in our galaxy do decay converting eventually into two or three very energetic $V$'s and some softer $V$'s which all  decay to standard model particles.

Without loss of generality, we imagine there is just one type of bound state. Then its number density $n_B$ satisfies the rate equation
\begin{equation}
\frac{d n_B}{d t} = - \Gamma_B n_B + \frac{1}{4} n_D^2 (\sigma v)_{\rm B} \ ,
\end{equation}
where $n_D$ is the unbound DM particle density \footnote{$n_D$ is the sum of DM and anti-DM particle number densities.}, $\Gamma_B$ is the DM decay rate and $(\sigma v)_{\rm B}$ is the DM bound state production cross section multiplied by the relative velocity of the DM particles. In the galaxy, if only a small fraction of DM have formed bound states or annihilated today, the free DM density $n_D$ is  constant in time. Solving the rate equation for the number density of dark bound states using this approximation  gives
\begin{equation}
n_B (t)= \frac{1}{4 \Gamma_B} n_D^2 (\sigma v)_{\rm B} \left( 1- e^{-\Gamma_B t} \right) \ .
\end{equation}
Taking the time $t$ to be the age of our galaxy $\tau_g \sim 10^{18}$ seconds, for $\Gamma_B t \gg1$, the value of $n_B$ approaches an equilibrium value given by,  
$4n_B\Gamma_B=n_D (\sigma v)_{\rm B}$.

The DM today includes  both free DM and the ones inside (unstable) bound states $n_D^{\rm (tot)} = n_D + 2 n_B$.
The total annihilation rate $R$ relevant for indirect detection signal, including DM annihilation both directly ($ \propto (\sigma v)_A$) and inside the bound states ($\propto (\sigma v)_B$), is
\begin{eqnarray}
R = 2 n_B \Gamma_B + \frac{1}{2} n_D^2 (\sigma v)_{\rm A} = \frac{1}{2} n_D^2 \left[ (\sigma v)_{\rm A} + (\sigma v)_{\rm B} \rule{0mm}{4mm}\right] \ ,
\end{eqnarray}
where in the second step, we have used the above equilibrium value for $n_B$. Hence the indirect detection signal discussed in the next section is fully determined by the sum of DM direct annihilation and bound state formation rates.

\section{Indirect Detection}\label{ID}

In this section, we quantify the importance of bound state formation on DM indirect detection.
The $V$ particles from the annihilation of DM will further decay into SM charged particles via the kinetic mixing with the photon and $Z$ boson, and contribute to the cosmic gamma ray spectrum. We will consider the Fermi constraint on the photon spectrum from DM annihilation at the galactic center~\cite{TheFermi-LAT:2015kwa}. 
The goal of this paper is not to provide a comprehensive list of all the constraints but rather to highlight the important role bound state formation plays in some regions of parameter space.

The gamma ray flux at the earth is obtained from  the DM annihilation rate averaged over the galactic center region via,
\begin{eqnarray}\label{spectrum3}
\frac{d \Phi_\gamma}{d E_\gamma} = \sum_{n=2}^3 \frac{1}{16\pi m_D^2} \frac{d N^{(n)}_\gamma}{d E_\gamma} \int d \Omega \int_{\rm l.o.s.} ds \rho(r(s, \theta))^2 (\sigma v)_{nV} \ ,
\end{eqnarray}
where $\rho$ is the DM density profile, $s$ is the distance of the annihilation point to the earth, $\theta$ is the angle between the line of sight and the direction of the galactic center in view of the earth, $r(s, \theta) =\sqrt{r_\odot^2 + s^2 - 2 r_\odot s \cos\theta}$, and $r_\odot = 8.4\,$kpc is the distance between the earth (or the sun) and the galactic center. We use the NFW profile for dark matte mass density distribution in the galaxy,
\begin{eqnarray}
\rho(r) = \frac{\rho_0}{(r/R)(1+r/R)^2} \ , 
\end{eqnarray}
where $G_N$ is the Newton's constant, $R=20\,$kpc, and we choose $\rho_0=0.34 \, {\rm GeV/cm^3}$ such that near the solar system $\rho(r_\odot)=0.4\,{\rm GeV/cm^3}$.
For the gamma ray observation by Fermi-LAT, the $\Omega$ integral covers the $15^\circ\times15^\circ$ region around the galactic center~\cite{TheFermi-LAT:2015kwa}.

As discussed in the previous section, the annihilation of DM could happen in two ways 
\begin{eqnarray}
&\chi \bar \chi \to 2 V\ ,& \nonumber \\
&\chi \bar \chi \to B \to n V& (n=2,3) \ . \nonumber
\end{eqnarray}
The first line is the usual direct annihilation with Sommerfeld enhancement, while the second line corresponds to having bound state formation and then DM annihilating within the bound states. The number of dark photons resulting from this annihilation, $n=2,3$, depends on the $C$-parity $C=(-1)^{\ell+S}$ of the bound state $B$ that decays via DM annihilation. 
In Eq.~(\ref{spectrum3}), the function ${d N^{(n)}_\gamma}/{d E_\gamma}$ is the photon spectrum at the source, depending on the number of $V$'s in the final state $(n=2,3)$. 
We describe the details of our calculation of ${d N^{(n=2,3)}_\gamma}/{d E_\gamma}$ in appendix~\ref{app1}.

\begin{figure}[h]
\centerline{\includegraphics[width=0.55\columnwidth]{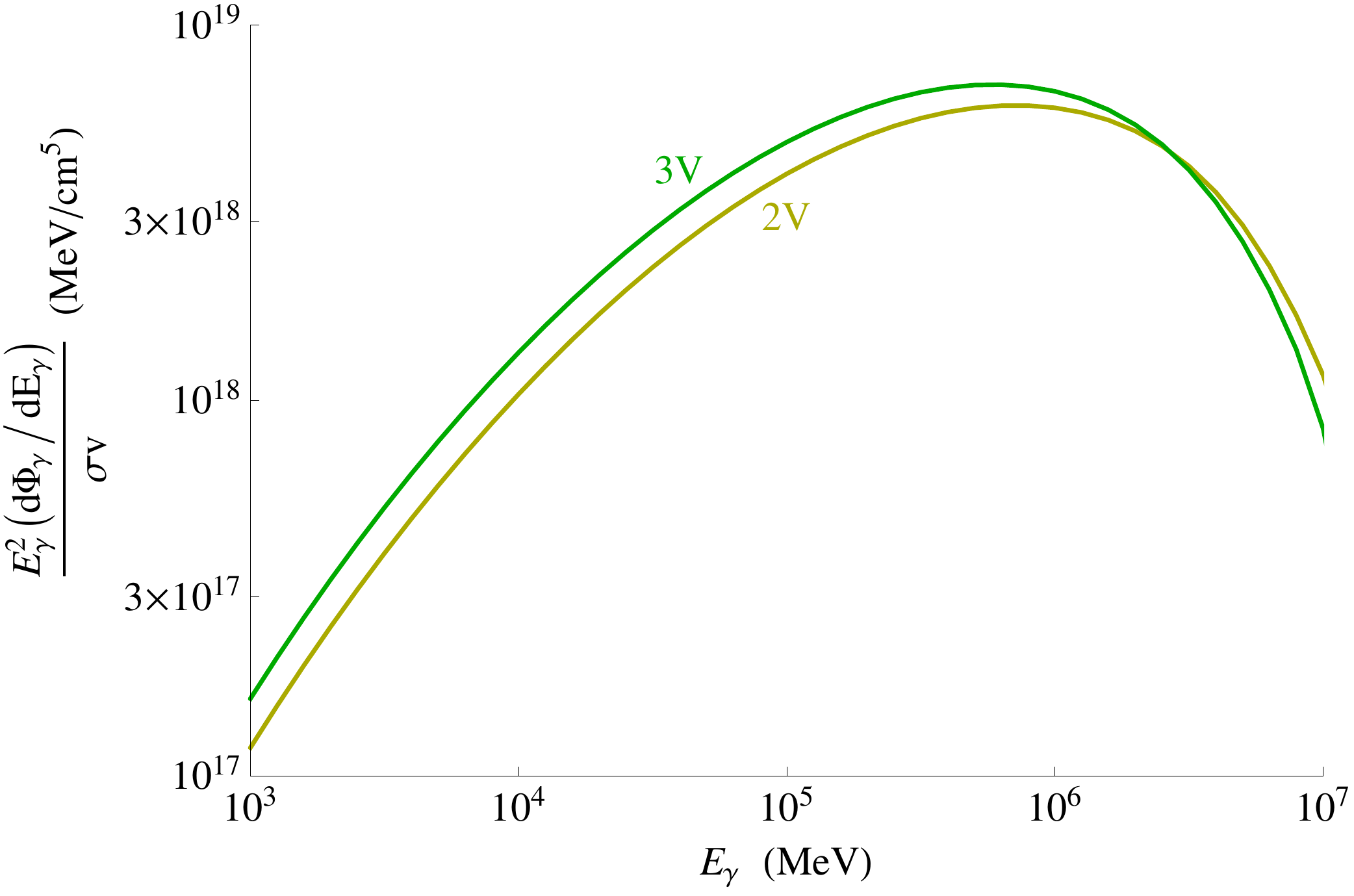}}
\caption{An example of gamma ray fluxes per unit cross section as a function of photon energy for the $2V$ and $3V$ channels.
Here the dark matter is $m_D=16.7\,{\rm TeV}$ and dark photon mass is $m_D=10\,{\rm GeV}$.
}\label{compareflux}
\end{figure}

The cross sections for $n=2,3$ are related to the ones discussed in the previous section as
\begin{align}
(\sigma v)_{2V} &=  (\sigma v)_{\rm A} + (\sigma v)_{\rm B} f_2 \ , \label{cross2}\\
(\sigma v)_{3V} &= (\sigma v)_{\rm B} (1-f_2) \ .\label{cross3}
\end{align}
The factor $f_2$ is the fraction of bound states annihilating into $2V$ from the state with $C=+1$. The rest of the bound states will annihilate to $3V$ with $C=-1$.
%Due to the spin multiplicity, bound states with even $\ell$ have 3 times the probability to annihilate decay into $3V$ compared with $2V$, and vice versa for odd $\ell$.
%But bound states could also de-excite.
%As long as the de-excitation with on-shell $V$ radiation is kinematically allowed, we find bound state with $n\lesssim1/\alpha_D$ (for any $\ell$) always prefers to first de-excite down to the ground state. Eventually, the annihilation decay of the ground state yields $f_2=0.25$.
%As another extreme, from the discussion in the previous section, bound state with both large $n$ and $\ell$ have to first de-excite down for certain levels, via off-shell $V$, before annihilation decay takes place. Because the annihilation decay into $2V$ is faster than into $3V$, in this case most of such states are likely to end up in $2V$, thus $f_2\simeq1$. Taking the two effects into account, the realistic value of $f_2$ should lie between 0.25 and 1.
As shown by Fig.~\ref{compareflux}, we find the $3V$ channel yields only a slightly larger (and slightly softer) gamma ray flux than the $2V$ channel.
Thus our numerical results are insensitive to the value of $f_2$.
For convenience, we use $f_2=1$ in the following calculations, which yields the most conservative limits.

In general, one has to calculate the cross section $(\sigma v)_{nV}$ within the line of sight integral because the bound state formation cross section is velocity dependent as discussed in Fig.~\ref{velocitydependence}, and the DM velocity near the galactic center depends on the position $r$.
For simplicity, we neglect the $r$ dependence and assume Maxwell-Boltzmann velocity distribution of DM with a reasonable root-mean-square value $v_{rms}= 150\,{\rm km/s}$ throughout the signal region at the galactic center as suggested in~\cite{Finkbeiner:2010sm}. This approximation does not affect the main point of our work.

\begin{figure}[t]
\centerline{\includegraphics[width=0.6\columnwidth]{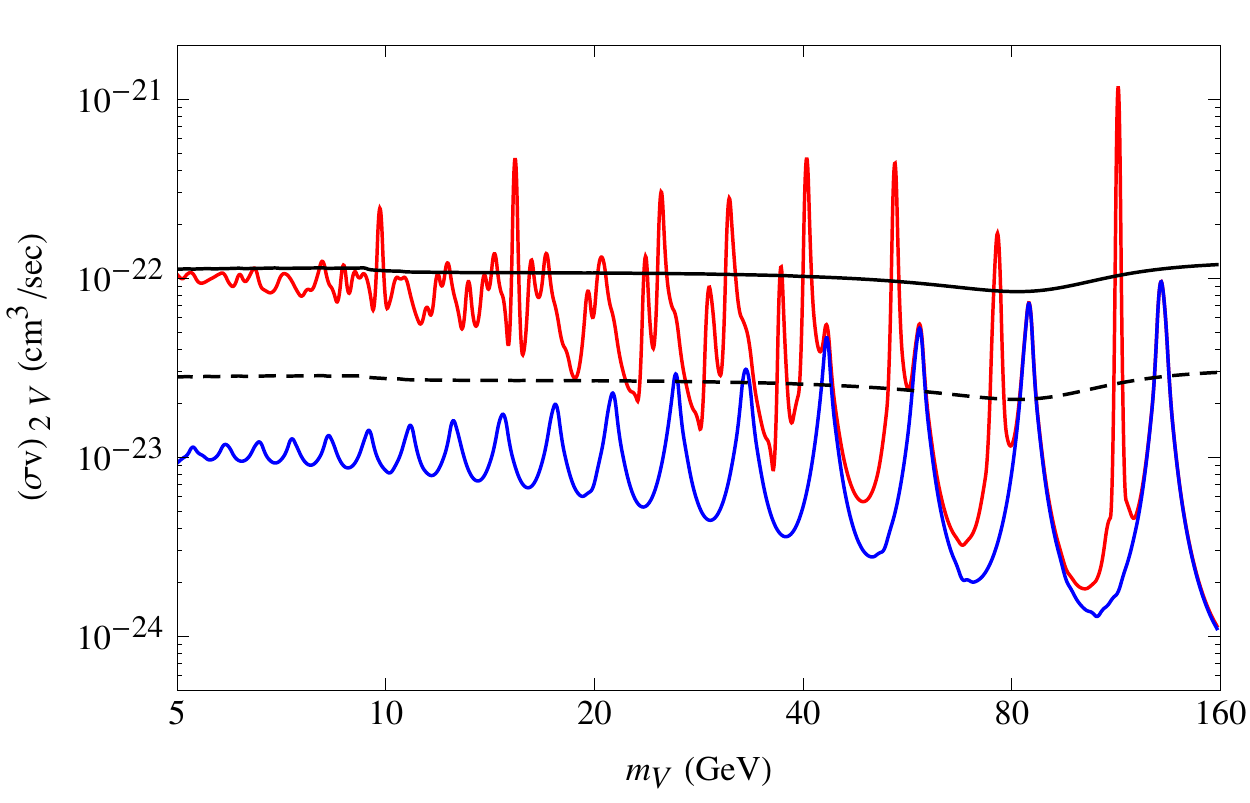}}
\caption{
Cross section for dark matter annihilation for indirect detection of gamma rays.
We take $m_D=16.7\,{\rm TeV}$ and $\alpha_D=0.2$.
The effect of dark matter bound states is included in the red curve (we choose $f_2=1$, see Eqs.~(\ref{cross2}) and (\ref{cross3})) but not the blue one.
The horizontal black curves correspond to the most conservative upper limit without including any interstellar emission background models (solid curve), 
and the the upper limit with background included (dashed curve).
Including the dark matter bound state formations results in a much stronger bound.
}\label{key}
\end{figure}

We compare the gamma ray spectrum with the one from the galactic center observed by Fermi-LAT~\cite{TheFermi-LAT:2015kwa}. We find that for the multi TeV DM in this study, the resulting gamma ray spectrum is peaked around a few hundred GeV to a TeV, while in the Fermi data, a spectrum decreasing with energy is provided only in the window 1-100 GeV. Therefore, the last bin with $E_\gamma \sim 80\,$GeV provides the strongest upper limit. 
As discussed above, we assume that all DM bound states annihilate decay into $2V$. The relevant cross section is just 
\begin{eqnarray}
(\sigma v)_{2V} = (\sigma v)_{\rm A} + (\sigma v)_{\rm B} \ .
\end{eqnarray}
This quantity as a function of $m_V$ is shown in Fig.~\ref{key}, for $m_D=16.7$\,TeV, $\alpha_D=0.2$. For $m_V$ less than the typical DM momentum at the galactic center, $(\sigma v)_{2V}$ has similar dependence on $m_V$ as Fig.~\ref{mVdependence}, because the DM annihilation via bound state formation gives the dominant effect for indirect detection. In contrast, we also show the direct annihilation cross section with Sommerfeld enhancement (often considered in the literature) in the blue curve, which can be lower than the total effective cross section by an order of magnitude. The solid black curve is the most conservative upper limit on $(\sigma v)_{\rm eff}$ by assuming zero background and requiring the signal from annihilation itself does not exceed the Fermi observation. The dashed black curve corresponds to taking into account the background using the interstellar emission models discussed in~\cite{TheFermi-LAT:2015kwa}, which sets a more stringent upper limit than the conservative one by a factor of $\sim$\,4.
Clearly, the effect of bound state formation at the galactic center can make a large difference. Given the other parameters, 
the upper limit on the dashed curve already rules out the region $m_V<20\,$GeV, while it would still be allowed if we only considered the direct annihilation channel with Sommerfeld enhancement.

\begin{figure}[t]
\centerline{\includegraphics[width=0.9\columnwidth]{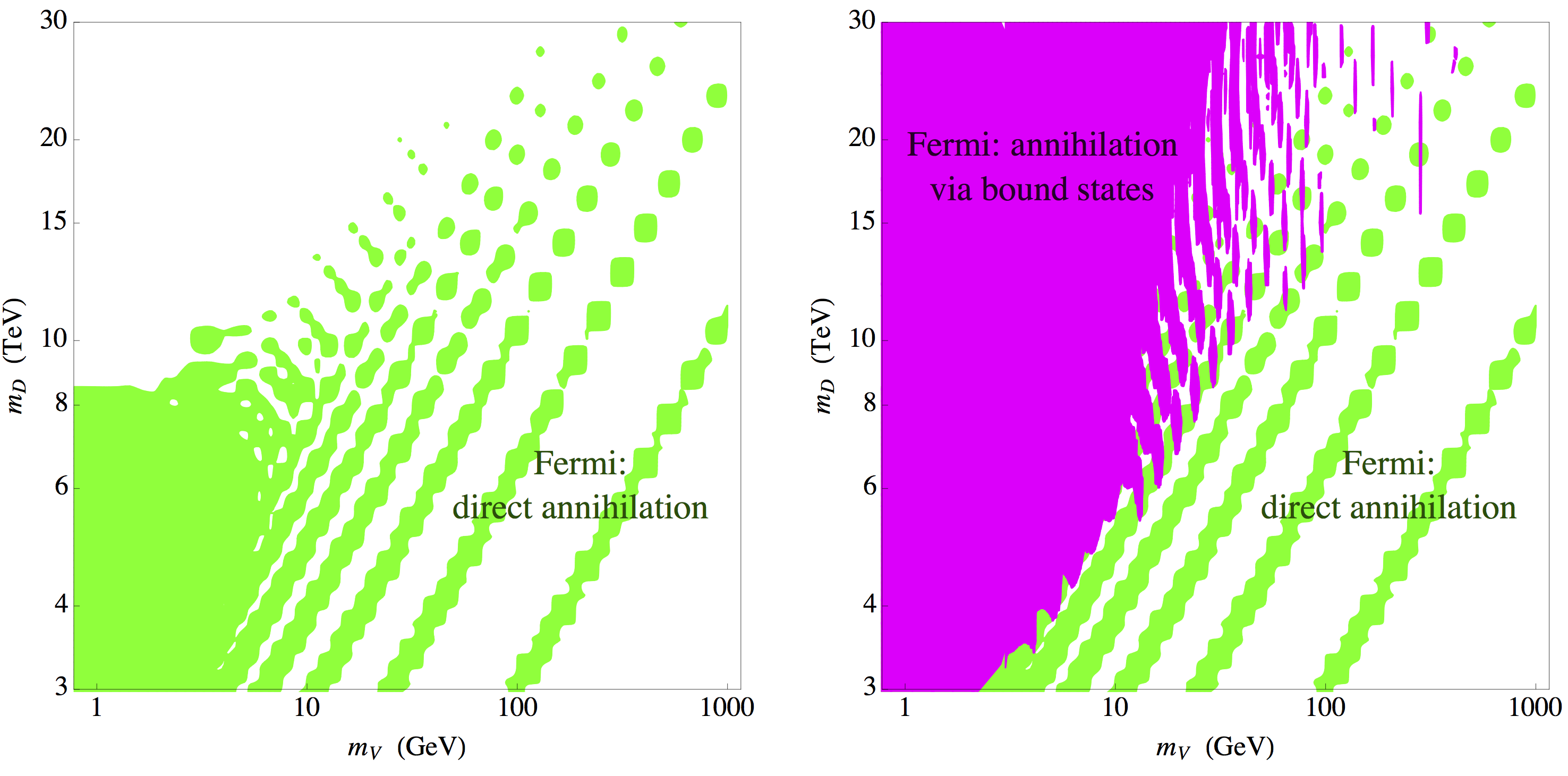}}
\caption{
The importance of dark matter bound states for constraining the parameter space of the dark matter--dark photon model. 
Here we zoom in to the region of parameter space marked by ``Focus of this study'' in Fig.~{parameterspace} in the $m_V-m_D$ plane. 
The green region shows the exclusion from indirect detection of gamma ray and includes only dark matter direct annihilation from the galactic center.
The magenta region in the plot on the right shows the (stronger) constraint when dark matter bound state formation effect is taken into account.
}\label{final}
\end{figure}

In Fig.~\ref{final}, we show the impact of bound state formation on indirect detection in the $m_V$ versus $m_D$ parameter space plane.
The value of $\alpha_D$ is chosen to give the correct thermal relic density for DM (see also the discussion in the next section).
In the plot on the left, we calculate the DM direct annihilation cross section with Sommerfeld enhancement and show the Fermi gamma ray constraint including the astrophysical background. 
The green region is excluded. 
In the plot on the right, we include the effect from DM bound state formation and the magenta region is further ruled out.
Clearly, bound state effects can play a very important role and must be included for DM masses above a few TeV.
In particular, for the window $m_V\in (1-10)\,$GeV, thermal DM with $m_D \gtrsim 8\,$TeV is allowed by the Fermi gamma ray data if we only consider the Sommerfeld enhanced direct annihilation. However, if we take into account of the contribution from the annihilation via bound state formation, 
the region of allowed DM mass increases to, $m_D > 30\,$TeV.

\section{Thermal Relic Density}\label{TRD}

In this section, we discuss the impact of bound state formation on DM thermal freeze out. As shown in Fig.~\ref{velocitydependence}, around the freeze out temperature when the DM velocity is  $v\sim0.3$, the bound state formation cross section is comparable to the direct annihilation one. One might think this would modify the effective annihilation cross section and in turn the value of $\alpha_D$ that gives rise to the observed thermal relic DM density. 

However, there is another important process which is bound state dissociation. Because the universe is hot during the time of freeze out, there is a plasma of the mediator $V$ particles around. For the part of the parameter space where bound states exist, the freeze out temperature $T_f$ is much larger than $m_V$. The dissociation process happens when an energetic $V$ particle collides with a bound state and breaks it into free $\chi$ and $\bar \chi$ particles. 
Because $\alpha_D/v \sim 1$ during freeze out, capture into the first few bound state energy levels dominate.
We will use the $S=0$ ground state to compare the dissociation and decay rates. \footnote{The $s=1$ state decays into $3V$ at a slower rate, and the $n>1$ bound states are shallower and easier to be dissociated.}

The decay rate of the ground state which, we call $\eta_D$, is approximately
\begin{eqnarray}
\Gamma(\eta_D\to2V) = \frac{1}{2} \alpha_D^5 m_D \ ,
\end{eqnarray}
where we neglect the impact of the $\eta_D$ binding energy on the energies of the final state $V$'s.

The dissociation rate of  $\eta_D$ is
\begin{eqnarray}
\Gamma(V\eta_D\to \chi \bar \chi) = \left[ \frac{3\zeta[3]}{\pi^2} T_f^3 \right] \left[ \frac{8\sqrt{3}}{9} \left(\frac{m_D}{T_f}\right)^3 \right]  \left[ \frac{128 \pi \alpha_D^5}{9 T_f^{1/2} m_D^{3/2} (\alpha_D^2 + 3T_f/m_D)}  \right] \ ,
\end{eqnarray}
where the first factor is the number density of $V$ particles, the second factor is the ratio of the bound state dissociation cross section to the formation one obtained in~\cite{Wise:2014jva} (which applies for both vector and scalar mediator cases), and the last factor is the bound state formation cross section Eq.~(\ref{KramersN}) for $n=1$. We have also used the condition $v \simeq \sqrt{3T_f/m_D}$ and the  approximation $T_f\gg m_D$.

Using the usual thermal value of $\alpha_D$ that gives the DM relic density, and the typical freeze out temperature $T_f \sim m_D/30$, we find the ratio
\begin{eqnarray}
\frac{\Gamma(V\eta_D\to \chi \bar \chi)}{\Gamma(\eta_D\to2V)} > 10 \ ,
\end{eqnarray}
for all the parameter space where $\alpha_D<0.3$. Therefore, the DM bound state is quickly dissociated by a collision with a $V$ before it has enough time to decay by DM-anti-DMDM annihilation. We conclude that  bound state formation is unimportant during the thermal freeze out of DM.

\section{CMB}\label{cmb}

Dark matter annihilation during recombination injects energy in the universe and could distort the CMB spectrum.
In this era, the DM velocity is very low, $v\sim 10^{-10} \ll m_V/m_D$ and $\alpha_D$.  One cannot take a very tiny dark photon mass to violate this condition, otherwise it would cause to too strong DM self-interactions and run into conflict with the bullet cluster observation~\cite{Randall:2007ph}.
We find that the bound state formation only involves the transition from $s$-wave scattering state to $p$-wave bound states, and its cross section is much lower than the direct annihilation one with the Sommerfeld enhancement (see Fig.~\ref{velocitydependence}).
Thus the usual constraint from CMB still applies~\cite{Galli:2009zc, Slatyer:2009yq, Slatyer:2015jla}.
For such low velocity, we use an approximate Sommerfeld factor $S$ obtained from the Hulth\'en potential~\cite{Cassel:2009wt, Feng:2010zp}, which is bounded from below,
$S \geq 6 \alpha_D m_D/m_V$.
We presented the CMB excluded region of parameter space in yellow in Fig.~\ref{parameterspace}.

\section{Summary and Discussion}\label{finale}

One of the simplest and most studied models of dark matter is a SM singlet Dirac fermion that annihilates down to its relic density through its coupling to a massive dark photon. We have shown that for dark matter mass in the tens of TeV range and dark photon mass in the GeV range, indirect detection constraints for dark matter in our galaxy are highly impacted by annihilation through dark-matter-anti-dark-matter bound states. In the regions where bound state formation is most important, annihilation through all possible bound states must be taken into account. In this work, we derived the general cross section for bound state formation with the radiation of a dark photon, and explored its dependence in the dark matter velocity and the dark photon mass.
Our most important results are illustrated by Fig.~\ref{final} where the magenta region shows the additional parameter space ruled out when annihilation through bound states is taken into account. 
The effects are strongest for large $\alpha_D$ and when the dark photon mass is smaller than the typical momentum of dark matter in the galaxy.
We have also argued that bound state effects are not important for dark matter annihilation during freeze out and recombination.

For dark matter indirect detection, we have only discussed the Fermi gamma ray constraint.
Our goal in this paper is to point out the importance of bound state formation rather than providing a comprehensive list of all the constraints. 
The bound state effects are expected to generic, and
a more complete analysis of the bound state effects on indirect detection via various cosmic ray components will be published elsewhere~\cite{elsewhere}.

Our results so far are based on the dark matter bound state formation cross section taking into account the emission of a single dark photon. In the limit of $m_V \rightarrow 0$, this cross section section is given by the Kramers formula in Eq.~(\ref{Kramers}). Its ratio to the s-wave geometric cross section $\sigma_G = 4\pi/k^2$ is 
\begin{equation}
\frac{\sigma_{\rm B}}{\sigma_G} \sim \alpha_D^3 \log\left(\frac{\alpha_D}{v}\right) \ .
\end{equation}
For the parameters of interest in this paper $\sigma_{\rm B}/\sigma_G \ll 1$, the unitarity bound is satisfied and hence we expect perturbation theory to be valid.

Finally, we comment on the case when the light mediator is a real scalar instead of a dark photon, which has also been considered in the literature. As pointed out in~\cite{Wise:2014jva}, because the operator for radiating a scalar is the unit operator and the scattering and bound state wave-functions are orthogonal, the leading order matrix element arises from second order in the multiple expansion. As a result, the bound state formation cross section would go as $\alpha_D^5$, in contrast to the $\alpha_D^3$ for the dark photon model. Thus, the effects of bound state formation on dark matter indirect detection is weaker in the scalar model.
It is worth mentioning that the scalar force is also attractive for same sign $\chi$'s and could result in bound states with a large number of DM particles that are stable, 
and may be cosmological important in the asymmetric dark matter case~\cite{Wise:2014ola}.

\section*{Acknowledgement} 

We thank Clifford Cheung, Walter D. Goldberger and  Maxim Pospelov for useful discussions. This work is supported by the DOE Grant DE-SC0011632, DE-FG02-92ER40701, DE-SC0010255, and by the Gordon and Betty Moore Foundation through Grant No.~776 to the Caltech Moore Center for Theoretical Cosmology and Physics. We are also grateful for the support provided by the Walter Burke Institute for Theoretical Physics. HA acknowledges the hospitality from the Perimeter Institute for Theoretical Physics.

\appendix

\section{Prompt photon spectrum}\label{app1}

For $V$ heavier than a few GeV perturbative methods for calculating $V$ decaying to quarks are applicable. We extrapolate those results into the region of lighter $V$ much of which is already strongly constrained by the CMB, see Fig.~\ref{parameterspace}. The prompt photons from the products of $V$ decay dominate the source term ${d N^{(n)}_\gamma}/{d E_\gamma}$ at large photon energy $E_\gamma$, and results in a peak in the spectrum. Other contributions to gamma ray from bremsstrahlung and inverse Compton scattering by charged particles (electrons) in the final states are only important for photon energy much lower than the peak energy~\cite{Meade:2009iu}.
We first calculate the energy spectrum of $V$ from dark matter annihilations.
Because the dark matter is non-relativistic in the galaxy, the energy distribution of $2V$ final state per reaction is always
\begin{eqnarray}
\frac{d N_V^{(2)}}{d y}(y) = 2 \delta(y - 1) \ ,
\end{eqnarray}
where $y=E_V/m_D$. On the other hand, for the $3V$ final state, the distribution is (for $m_V\ll m_D$)~\cite{An:2015pva}
\begin{eqnarray}
\frac{d N_V^{(3)}}{d y}(y) = \frac{9}{4 (\pi^2-9) y^2} \left[ y(3y-8) + \frac{(y-1)(y^2 -6y +16) \ln(1-y)}{y-2} \right] \left({1 \over 1- m_V/m_D- 3 m_V^2/(4 m_D^2)} \right) ,
\end{eqnarray}
and in this case  $y$ is between $y_{\rm min} = {m_V}/{m_D}$, $y_{\rm max} = 1- {3 m_V^2}/({4 m_D^2})$.
The $V$'s will subsequently decay into charged fermion pairs $f$, $\bar f$. It is easy to show that the energy distribution of $f$ in the boosted $V$ frame is flat,
\begin{eqnarray}
\frac{d N_f}{d E_f} (y) \simeq \frac{2}{y m_D \sqrt{1- \frac{4 m_f^2}{m_V^2}}} \ , \ \ \
(E_f)_{\rm min}^{\rm max}(y) = \frac{1}{2} y m_D \left( 1 \pm \sqrt{1- \frac{4 m_f^2}{m_V^2}} \right) \ .
\end{eqnarray}
Here we have used the approximation that for $m_D \gg m_V$, in most of the final state phase space $E_V = y m_D \gg m_V$.
Therefore, the photon energy spectrum (prompt, from final state radiation) per annihilation is given by
\begin{eqnarray}
\frac{d N^{(n, f)}_\gamma}{d E_\gamma} = \int_{y_{\rm min}}^{y_{\rm max}} dy\ \frac{d N_V^{(n)}}{d y}(y)
\int_{(E_f)_{\rm min}(y)}^{(E_f)_{\rm max}(y)} d E_f\ \frac{d N_f}{d E_f}(y) \ \frac{d N^0_{\gamma}}{d E_\gamma}(E_f, E_\gamma/E_f) \ ,
\end{eqnarray}
where $d N^0_{\gamma}/d E_\gamma$ is the photon distribution function per each injection of a DM charged particle $f$, and is obtained using the code {\sf PPPC4}~\cite{Cirelli:2010xx}. 
Finally, the total photon energy spectrum is obtained by summing over the possible fermion flavors $f$, weighed by the branching ratio of the decay $V\to f \bar f$,
\begin{eqnarray}\label{spectrum}
\frac{d N^{(n)}_\gamma}{d E_\gamma} = \sum_f  \frac{d N^{(n, f)}_\gamma}{d E_\gamma} {\rm Br}(V\to f \bar f) \ .
\end{eqnarray}
We give the decays rates for $V\to f \bar f$ in  appendix~\ref{Br}.

For the $3V$ channel ($n=3$) Eq.~(\ref{spectrum}) can be simplified by interchanging  orders of integration. After some algebra,
\begin{eqnarray}\label{spectrum2}
\frac{d N^{(3,f)}_\gamma}{d E_\gamma} = 
\int_{(E_f)_{\rm min}(y_{\rm min})}^{(E_f)_{\rm max}(y_{\rm max})} d E_f 
\frac{1}{m_D} \left[ F\left( y_{+} \right)- F\left( y_{-} \right) \rule{0mm}{4mm}\right] \
\frac{d N_{\gamma}}{d E_\gamma}(E_f, E_\gamma/E_f) \ ,
\end{eqnarray}
where 
\begin{eqnarray}
F(y) = \frac{-9 \left[ 4 y + \ln(1-y) (4 - 7 y + 3 y^2 - y^2 \ln(2-y)) - y^2 {\rm PolyLog}(2, y-1) \right]}{2 y^2 (\pi^2 -9) 
\left (1- {m_V}/{m_D} - {3 m_V^2}/({4 m_D^2}) \right)  \sqrt{1 - 4 m_f^2 / m_V^2} } \ ,
\end{eqnarray}
and 
\begin{eqnarray}
y_+ = {\rm Min} \left[ 1-\frac{3 m_V^2}{4 m_D^2}, \ \frac{2 E_f/m_D}{1- \sqrt{1- \frac{4 m_f^2}{m_V^2}}} \right], \ \ \
y_- = {\rm Max} \left[ \frac{m_V}{m_D}, \ \frac{2 E_f/m_D}{1+ \sqrt{1- \frac{4 m_f^2}{m_V^2}}} \right] \ .
\end{eqnarray}

\section{Dark photon decay rates}\label{Br}

In general, the kinetic mixing between the dark photon and the usual photon originates from the $SU(2)\times U(1)$ gauge invariant operator $\frac{\kappa}{2\cos\theta_w} B_{\mu\nu} V^{\mu\nu}$, where $B_\mu$ is the gauge field for hypercharge. After the electroweak symmetry breaking, this operator not only induce the kinetic mixing term 
$\frac{\kappa}{2} F_{\mu\nu} V^{\mu\nu}$ in Eq. (1), but also a kinetic mixing between $V$ and the $Z$ boson, $-\frac{\kappa\tan\theta_w}{2} Z_{\mu\nu} V^{\mu\nu}$. Therefore, the branching ratios of $V$ is not simply like those of a massive photon.
The kinetic mixing between $V$ and $Z$ can play an important role for $m_V\gg$\,GeV. 

\begin{figure}[h]
\centerline{\includegraphics[width=0.55\columnwidth]{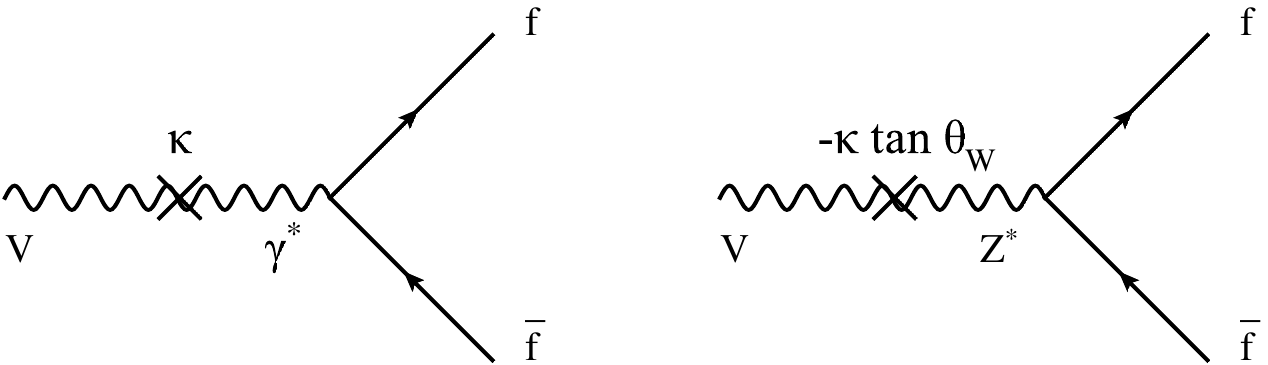}}
\caption{Feynman diagrams for the dark photon $V$ decaying to fermions, via off-shell photon or $Z$-boson.}\label{Feynman}
\end{figure}

The Feynman diagrams for $V$ decaying to fermion pairs via off-shell photon and $Z$-boson are shown in Fig.~\ref{Feynman}. The effective coupling of an on-shell $V$ to the left- (right-)handed fermion current is
\begin{eqnarray}
g_L &=& \kappa g \sin\theta_w \left[ Q_f - \frac{m_V^2}{m_V^2 - m_Z^2} \frac{1}{\cos^2\theta_w} (T_{3}^f - Q_f \sin^2\theta_w) \right] \ , \nonumber \\
g_R &=& \kappa g \sin\theta_w \left[ Q_f - \frac{m_V^2}{m_V^2 - m_Z^2} \frac{1}{\cos^2\theta_w} ( - Q_f \sin^2\theta_w) \right] \ .
\end{eqnarray}
The decay rates are then~\cite{Curtin:2014cca}
\begin{eqnarray}
\Gamma_{V\to f \bar f} = \frac{N_c^f m_V}{24\pi} (1-4r_f)^{1/2} \left[ g_L^2 (2 - 2 r_f) + g_R^2 ( 2- 2 r_f) - 12 g_L g_R r_f \right] \ ,
\end{eqnarray}
where $r_f = m_f^2/m_V^2$, and $N_c^f=3$ for quarks and 1 for charged leptons and neutrinos. For the parameter space of interest to this study, $m_V$ lies between $\sim {\rm GeV}$ and the weak scale. We obtain the total decay rate by summing over all possible quark and lepton flavors that are kinematically allowed for $V$ to decay into.

\end{document}